
%
%
\newdimen\FigSize	\FigSize=.9\hsize 
%
\newskip\abovefigskip	\newskip\belowfigskip
\gdef\epsfig#1;#2;{\par\vskip\abovefigskip\penalty -500
   {\everypar={}\epsfxsize=#1\noindent
    \centerline{\epsfbox{#2}}}%
    \vskip\belowfigskip}%
%
\newskip\figtitleskip
\gdef\tepsfig#1;#2;#3{\par\vskip\abovefigskip\penalty -500
   {\everypar={}\epsfxsize=#1\noindent
    \vbox
      {\centerline{\epsfbox{#2}}\vskip\figtitleskip
       \centerline{\figtitlefont#3}}}%
    \vskip\belowfigskip}%
%
\newcount\FigNr	\global\FigNr=0
\gdef\nepsfig#1;#2;#3{\global\advance\FigNr by 1
   \tepsfig#1;#2;{Figure\space\the\FigNr.\space#3}}%
%
%
%
\gdef\ipsfig#1;#2;{
   \midinsert{\everypar={}\epsfxsize=#1\noindent
	      \centerline{\epsfbox{#2}}}%
   \endinsert}%
%
\gdef\tipsfig#1;#2;#3{\midinsert
   {\everypar={}\epsfxsize=#1\noindent
    \vbox{\centerline{\epsfbox{#2}}%
          \vskip\figtitleskip
          \centerline{\figtitlefont#3}}}\endinsert}%
%
\gdef\nipsfig#1;#2;#3{\global\advance\FigNr by1%
  \tipsfig#1;#2;{Figure\space\the\FigNr.\space#3}}%
\newread\epsffilein    
\newif\ifepsffileok    
\newif\ifepsfbbfound   
\newif\ifepsfverbose   
\newdimen\epsfxsize    
\newdimen\epsfysize    
\newdimen\epsftsize    
\newdimen\epsfrsize    
\newdimen\epsftmp      
\newdimen\pspoints     
\pspoints=1bp          
\epsfxsize=0pt         
\epsfysize=0pt         
\def\epsfbox#1{\global\def\epsfllx{72}\global\def\epsflly{72}%
   \global\def\epsfurx{540}\global\def\epsfury{720}%
   \def\lbracket{[}\def\testit{#1}\ifx\testit\lbracket
   \let\next=\epsfgetlitbb\else\let\next=\epsfnormal\fi\next{#1}}%
\def\epsfgetlitbb#1#2 #3 #4 #5]#6{\epsfgrab #2 #3 #4 #5 .\\%
   \epsfsetgraph{#6}}%
\def\epsfnormal#1{\epsfgetbb{#1}\epsfsetgraph{#1}}%
\def\epsfgetbb#1{%
%
%
\openin\epsffilein=#1
\ifeof\epsffilein\errmessage{I couldn't open #1, will ignore it}\else
%
%
   {\epsffileoktrue \chardef\other=12
    \def\do##1{\catcode`##1=\other}\dospecials \catcode`\ =10
    \loop
       \read\epsffilein to \epsffileline
       \ifeof\epsffilein\epsffileokfalse\else
%
%
          \expandafter\epsfaux\epsffileline:. \\%
       \fi
   \ifepsffileok\repeat
   \ifepsfbbfound\else
    \ifepsfverbose\message{No bounding box comment in #1; using defaults}\fi\fi
   }\closein\epsffilein\fi}%
%
%
\def\epsfsetgraph#1{%
   \epsfrsize=\epsfury\pspoints
   \advance\epsfrsize by-\epsflly\pspoints
   \epsftsize=\epsfurx\pspoints
   \advance\epsftsize by-\epsfllx\pspoints
%
%
   \epsfxsize\epsfsize\epsftsize\epsfrsize
   \ifnum\epsfxsize=0 \ifnum\epsfysize=0
      \epsfxsize=\epsftsize \epsfysize=\epsfrsize
%
%
     \else\epsftmp=\epsftsize \divide\epsftmp\epsfrsize
       \epsfxsize=\epsfysize \multiply\epsfxsize\epsftmp
       \multiply\epsftmp\epsfrsize \advance\epsftsize-\epsftmp
       \epsftmp=\epsfysize
       \loop \advance\epsftsize\epsftsize \divide\epsftmp 2
       \ifnum\epsftmp>0
          \ifnum\epsftsize<\epsfrsize\else
             \advance\epsftsize-\epsfrsize \advance\epsfxsize\epsftmp \fi
       \repeat
     \fi
   \else\epsftmp=\epsfrsize \divide\epsftmp\epsftsize
     \epsfysize=\epsfxsize \multiply\epsfysize\epsftmp   
     \multiply\epsftmp\epsftsize \advance\epsfrsize-\epsftmp
     \epsftmp=\epsfxsize
     \loop \advance\epsfrsize\epsfrsize \divide\epsftmp 2
     \ifnum\epsftmp>0
        \ifnum\epsfrsize<\epsftsize\else
           \advance\epsfrsize-\epsftsize \advance\epsfysize\epsftmp \fi
     \repeat     
   \fi
%
%
   \ifepsfverbose\message{#1: width=\the\epsfxsize, height=\the\epsfysize}\fi
   \epsftmp=10\epsfxsize \divide\epsftmp\pspoints
   \vbox to\epsfysize{\vfil\hbox to\epsfxsize{%
      \includegraphics{#1}%
      \hfil}}%
\epsfxsize=0pt\epsfysize=0pt}%
%
%
{\catcode`\%=12 \global\let\epsfpercent=
%
%
\long\def\epsfaux#1#2:#3\\{\ifx#1\epsfpercent
   \def\testit{#2}\ifx\testit\epsfbblit
      \epsfgrab #3 . . . \\%
      \epsffileokfalse
      \global\epsfbbfoundtrue
   \fi\else\ifx#1\par\else\epsffileokfalse\fi\fi}%
%
%
\def\epsfgrab #1 #2 #3 #4 #5\\{%
   \global\def\epsfllx{#1}\ifx\epsfllx\empty
      \epsfgrab #2 #3 #4 #5 .\\\else
   \global\def\epsflly{#2}%
   \global\def\epsfurx{#3}\global\def\epsfury{#4}\fi}%
%
%
\def\epsfsize#1#2{\epsfxsize}%
%
%

\epsfverbosetrue			
\abovefigskip=\baselineskip		
\belowfigskip=\baselineskip		
\global\let\figtitlefont\bf		
\global\figtitleskip=.5\baselineskip	
\magnification=\magstep1
\pageno=0
\font\huge=cmbx10 scaled\magstep2
\font\xbold=cmbx10 scaled\magstep1

\font\sub=cmbx10
\font\small=cmr7

 at 10truept
\count5=0
\count6=1
\count7=1
\count8=1

\def\references{\bigskip\noindent{\bf References}\hfill\break}

\def\abstract#1\par{\noindent{\sub Abstract.} #1 \par}
\def\equ(#1){\hskip-0.03em\csname e#1\endcsname}
\def\clm(#1){\csname c#1\endcsname}
\def\equation(#1){\eqno\tag(#1)}
\def\tag(#1){(\number\count5.
	      \number\count6)
    \expandafter\xdef\csname e#1\endcsname{
    (\number\count5.\number\count6)}
    \global\advance\count6 by 1}
\def\claim #1(#2) #3\par{
    \vskip.1in\medbreak\noindent
    {\bf #1\ \number\count5.\number\count7.\ }{\sl #3}\par
    \expandafter\xdef\csname c#2\endcsname{#1~\number\count5.\number\count7}
    \global\advance\count7 by 1
    \ifdim\lastskip<\medskipamount
    \removelastskip\penalty55\medskip\fi}
\def\section#1\par{\vskip0pt plus.3\vsize\penalty-75
    \vskip0pt plus -.3\vsize\bigskip\bigskip
    \global\advance\count5 by 1
    \message{#1}\leftline
     {\xbold \number\count5.\ #1}
    \count6=1
    \count7=1
    \count8=1
    \nobreak\smallskip\noindent}
\def\subsection#1\par{\vskip0pt plus.2\vsize\penalty-75
    \vskip0pt plus -.2\vsize\bigskip\bigskip
    \message{#1}\leftline{\sub
    \number\count5.\number\count8.\ #1}
    \global\advance\count8 by 1
    \nobreak\smallskip\noindent}
\def\proofof(#1){\medskip\noindent{\bf Proof of \csname c#1\endcsname.\ }}
\footline{\ifnum\pageno=0\hss\else\hss\tenrm\folio\hss\fi}

\let\cl=\centerline

\let\wt=\widetilde
\let\eps=\varepsilon
\let\sss=\scriptscriptstyle
\input amssym.def
\input amssym.tex
\def\RG{\Re}

\def\natural{{\Bbb N}}
\def\integer{{\Bbb Z}}

\def\real{{\Bbb R}}

\def\torus{{\Bbb T}}

\def\Id{{\Bbb I}}
\def\id{{\rm I}}
\def\half{{1\over 2}}

\def\CC{{\cal C}}

\def\JJ{{\cal J}}

\def\OO{{\cal O}}
\def\PP{{\cal P}}

\def\RR{{\cal R}}
\def\SS{{\cal S}}

\def\UU{{\cal U}}

\def\WW{{\cal W}}


\def\rED{1}
\def\rMEsc{2}
\def\rMcKii{3}
\def\rKi{4}
\def\rKii{5}
\def\rCGJi{6}
\def\rCGJii{7}
\def\rCGJK{8}
\def\rAK{9}
\def\rKad{10}
\def\rSK{11}
\def\rMcKi{12}
\def\rMcKiii{13}
\def\rSti{14}
\def\rMH{15}
\def\rACS{16}
\def\rMMS{17}
\def\rTh{18}
\def\rRdL{19}

\cl{{\huge A Renormalization Group for Hamiltonians:}}
\cl{{\huge Numerical Results}}
\vskip.6in
\cl{Juan J.~Abad, Hans Koch
\footnote{$^1$}
{{\small Supported in Part by the
National Science Foundation under Grant No. DMS--9705095.}}}
\cl{Department of Mathematics, University of Texas at Austin}
\cl{Austin, TX 78712}
\epsfbox{X.ps}
\vskip.25in
\cl{Peter Wittwer
\footnote{$^2$}
{{\small Supported in Part by the
Swiss National Science Foundation.}}}
\cl{D\'epartement de Physique Th\'eorique, Universit\'e de Gen\`eve}
\cl{Gen\`eve, CH 1211}
\vskip.8in
\abstract
We describe a renormalization group transformation
that is related to the breakup of golden invariant tori
in Hamiltonian systems with two degrees of freedom.
This transformation applies to a large class of Hamiltonians,
is conceptually simple,
and allows for accurate numerical computations.
In a numerical implementation, we find a nontrivial fixed point
and determine the corresponding critical index and scaling.
Our computed values for various universal constants
are in good agreement with existing data for area--preserving maps.
We also discuss the flow associated with the nontrivial fixed point.

\par\vfill\eject

\section Introduction

The renormalization group transformation considered in this paper
acts on analytic Hamiltonians for systems of two degrees of freedom.
It is designed to apply to the stability problem
for invariant tori associated with frequency vectors
$\omega=(\omega_1,\omega_2)$, for which
$\omega_1/\omega_2$ is a reduced quadratic irrational.
For simplicity, we will focus from the beginning
on the case of the golden mean.
Our formalism can also be extended to higher dimensions,
as in [\rKi], but the results may be qualitatively different;
see also [\rMH--\rMMS].

We start with some simple facts about the golden mean
$\vartheta=\half+\half\sqrt{5}$. The numbers
$\vartheta$ and $-1/\vartheta$ are solutions of a quadratic equation
with integer coefficients, namely the equation
for the eigenvalues of the matrix $T=\left[{0~1\atop 1~1}\right]$.
The corresponding eigenvectors are 
$\omega=(1/\vartheta,1)$ and $\Omega=(1,-1/\vartheta)$, respectively.
The matrix $T$ can also be used to generate
the continued fraction approximants $\vartheta_k=b_k/a_k$ of $\vartheta$.
More specifically, we have $(a_k,b_k)=T^k(1,1)$,
for $k=0,1,2,\ldots$.

The Hamiltonians considered here are assumed to have
a rapidly convergent Fourier--Taylor series of the form
$$
H(q,p)=\sum_{(\nu,\alpha)\in I}H_{\nu,\alpha}\,F_{\nu,\alpha}(q,p)\,,
\qquad F_{\nu,\alpha}(q,p)=
(\omega\cdot p)^{\alpha_1}(\Omega\cdot p)^{\alpha_2} e^{i\nu\cdot q} \,,
\equation(FTDef1)
$$
for all $q$ in $\real^2$, and for $p$ in some open neighborhood
of the origin in $\real^2$. The sum in this equation ranges over all
multi--indices $(\nu,\alpha)$ in $I=\integer^2\!\times\natural^2$,
with $\natural$ being the set of all non--negative integers.
We will refer to $p$ and $q$ as the action and (lifted) angle
variable, respectively. Recall that the flow defined by $H$ is 
$(\dot q,\dot p)=\bigl(\nabla_2 H,-\nabla_1 H\bigr)$,
where $\nabla_j$ denotes the gradient with respect to the $j$--th variable.

In order to motivate our analysis,
let us first consider the integrable Hamiltonian
$H_0(q,p)=\omega\cdot p+\half(\Omega\cdot p)^2$.
When restricted to a fixed level set $H_0(q,p)=E$,
this Hamiltonian has an invariant torus $\OO_\infty$
whose frequency vector is proportional (in fact equal) to $\omega$.
In addition, for every sufficiently large integer $k$,
we can find a periodic orbit $\OO_k$ whose frequency vector
is a constant multiple of $(a_k,b_k)$.
The same can be observed (and proved) for analytic Hamiltonians
that are sufficiently close to $H_0$.
In fact, the orbits $\OO_k$ accumulate at the torus $\OO_\infty\,$.
To give a more precise description,
assume for simplicity that each of the orbits $\OO_k$ contains
a unique point $(q_k,p_k)$ with $q_k=0$.
This holds e.g. for many Hamiltonians that satisfy $H(q,p)=H(-q,p)$.
Then $p_k$ approaches a limit $p_\infty$
as $k$ tends to infinity,
and the distance $d_k$ between $p_k$ and $p_\infty$ tends to zero
with an asymptotic ratio $d_{k+1}/d_k\to\vartheta^{-2}$.
This should not bee too astonishing, as
$|\vartheta_k-\vartheta|$ tends to zero with the same asymptotic ratio.
What is more astonishing is what seems to happen
in many one--parameter families
$\beta\mapsto H_0+\beta F$,
as the strength of the perturbation is increased:
By looking for orbits with a given degree of stability (e.g. residue $1$),
among the orbits $\OO_k$ for sufficiently large $k$,
one finds a convergent sequence $\beta_m,\beta_{m+1},\ldots$
in some parameter interval $J$, with the following property.
The Hamiltonian $H_0+\beta F$, with $\beta\in J$,
has an orbit with the given stability only if
$\beta=\beta_k$ for some $k\ge m$,
and in this case, the orbit is unique and equal to $\OO_k\,$.
Furthermore, the ratios $(\beta_{k-1}-\beta_k)/(\beta_k-\beta_{k+1})$
converge to a limit $\delta$, as $k$ tends to infinity;
and the ratios $d_{k+1}/d_k$, for the Hamiltonian 
with the parameter value $\beta_\infty=\lim_{k\to\infty}\beta_k$,
converge to a limit $\lambda$ different from $\vartheta^{-2}$.
The critical index $\delta$ and the critical scaling $\lambda$
appear to be independent of the family considered,
within a large class of Hamiltonians.
The observed values are $\delta=1.627\ldots$ and $\lambda=0.326\ldots$.
A more detailed description of these phenomena
can be found in References [\rED,\rMEsc,\rKad--\rMcKiii].

Such scaling and universality properties
suggest the use of renormalization group methods.
The idea is to find a transformation $\RR$,
acting on a space of Hamiltonians, which contracts differences
that are irrelevant for the phenomena being considered.
According to the description above, $\RR$ should therefore
have two fixed points:
One that is a prototype of a Hamiltonian with
trivial asymptotic scaling, attracting all other Hamiltonians of this type,
and another one that plays the same role
for Hamiltonians with critical scaling.
The fixed point equation $\RR(H)=H$ should of course imply
self--similarity for the set of $\vartheta_k$--orbits of $H$.
A transformation $\RG$ which has the potential of
satisfying all of the necessary requirements
(modulo the existence of a trivial unstable direction)
has been described in [\rKi]. It is of the following form:
$$
\RG(H)={\tau\over\mu}H\circ\UU_H\circ T_\mu -\epsilon\,,
\qquad T_\mu(q,p)=\bigl(Tq,\mu T^{-1}p\bigr)\,.
\equation(RGDef1)
$$
 $\UU_H$ is a canonical change of coordinates that will be
defined later; in particular, $\UU_H$ is the identity map
if $H(q,p)$ does not depend on $q$.
The constants $\epsilon=\epsilon(H)$,
$\tau=\tau(H)$, and $\mu=\mu(H)$ are determined by
three normalization conditions for $\wt H=\RG(H)$.
In this paper, we choose these conditions to be $\wt H_{0,0}=0$,
$\wt H_{0,(1,0)}=1$, and $\wt H_{0,(0,2)}=\half$.
If they cannot be satisfied, then $\RG(H)$ remains undefined.
It is easy to check that the integrable Hamiltonian
$H_0(q,p)=\omega\cdot p+\half(\Omega\cdot p)^2$
is a fixed point for $\RG$, with $\epsilon(H_0)=0$,
$\tau(H_0)=\vartheta$, and $\mu(H_0)=\vartheta^{-3}$.

The first step $H\mapsto H\circ\UU_H$ in the definition of $\RG(H)$
is similar in spirit to the averaging operation
used in KAM theory [\rTh,\rRdL]. Its purpose is to eliminate
irrelevant (non--resonant) degrees of freedom from $H$.
The next step $H\mapsto H\circ T_1$ represents a ``shift''
of frequencies, in the sense that
if $H$ has a periodic orbit for a frequency vector $\nu$,
then $H\circ T_1$ has a (trivially) related orbit
for the frequency vector $T^{-1}\nu$.
Notice that $T_1$ is a canonical transformation.
In order to re--normalize the resulting Hamiltonian,
we perform a rescaling $H\mapsto\mu^{-1}H(.,\mu .)$
of the action variable $p$,
a rescaling $H\mapsto\tau H$ of the energy (or time),
and subtract a trivial constant $\epsilon$.
We note that the shift and scalings are directly related
to the observation of self--similarity near invariant
$\omega$--tori. They have also been used in other
renormalization transformation [\rED--\rCGJK].

The change of coordinates $\UU_H$ will be chosen
within a restricted class of canonical transformations.
Each transformation $U: (q,p)\mapsto (q+Q,p+P)$
in this class is associated with a generating function $\phi(q,p)$
that is $2\pi$--periodic in the components of $q$,
and $U=U_\phi$ is defined implicitly by the equation
$$
\eqalign{
Q(q,p)&=\nabla_2\phi\bigl(q,p+P(q,p)\bigr)\,,\cr
P(q,p)&=-\nabla_1\phi\bigl(q,p+P(q,p)\bigr)\,.\cr}
\equation(GenFun1)
$$
One of the consequences of this restriction is that $\UU_H$
cannot include $p$--translations like $V_s: (q,p)\mapsto(q,p-s\Omega)$.
As a result, $\RG$ has a trivial unstable direction.
This could be avoided easily by including suitable translations
in the definition of our renormalization group transformation.
More specifically, we can define
$\RR(H)=\bigl(\RG(H\circ V_s)\bigr)\circ V_{-s}\,$,
with $s=s(H)$ determined in such a way that
$\wt H=\RR(H)$ satisfies $\wt H_{0,(0,1)}=0$.
Our numerical result for $\RG$, described in the next section,
can be translated into corresponding results for $\RR$.
They indicate that
$\RR$ has a trivial attractive fixed point (namely $H_0$),
and a nontrivial fixed point $H_\ast$
whose stable manifold is of codimension one.

Concerning the notion of relevant or irrelevant degrees of freedom
for analytic Hamiltonians,
we note that even a critical $\omega$--torus is smooth
in the direction of the flow.
This suggests that the most relevant perturbations
are the ones that distort the $\omega$--torus,
and the geometry of nearby orbits,
mainly in the directions transversal to the flow.
If we assume that (or work in coordinates where)
the average velocity on the $\omega$--torus of $H$
is proportional to $\omega$, then these perturbations
are given by the functions $F_{\nu,\alpha}$
whose frequency vectors $\nu$ are almost perpendicular to $\omega$.
They are usually referred to as ``resonant'' modes.
The idea is to eliminate all but these resonant modes.
Another way of seeing why this should be done is the following:
Assume that, as a function of the angle variable,
$H$ is analytic in a strip given by
$|{\rm Im~}\omega\cdot q|<r$ and $|{\rm Im~}\Omega\cdot q|<R$.
Then $H\circ T_1$ is analytic in a similar strip, but with
$r$ and $R$ replaced by $r\vartheta$ and $R/\vartheta$, respectively.
Thus, in order to avoid that $\RG$ shrinks the domain of analyticity,
we have to regularize $H$ in the direction $({\rm Im~}q)\propto\omega$.
One way of doing this is by eliminating the non--resonant modes of $H$.
To be more precise, we will call $F_{\nu,\alpha}$ a
resonant mode if $(\nu,\alpha)$ belongs to the set
$$
I^{+}=\bigl\{(\nu,\alpha)\in\integer^2\!\times\natural^2:
|\omega\cdot\nu|\le\sigma\|\nu\|+\kappa\|\alpha\|\bigr\}\,,
\equation(IminusDef1)
$$
or a non--resonant mode if $(\nu,\alpha)$ belongs to the
complement $I^{-}$ of this set in $\integer^2\!\times\natural^2$.
Here, $\sigma$ and $\kappa$ are positive constants to be chosen later,
and $\|.\|$ denotes the $\ell^1$ norm on $\real^2$,
that is, $\|x\|=|x_1|+|x_2|$.
Correspondingly,
$$
\Id^{\pm}H=\sum_{(\nu,\alpha)\in I^{\pm}}H_{\nu,\alpha}F_{\nu,\alpha}
\equation(IIminusDef1)
$$
will be referred to as the resonant ($+$) and non--resonant ($-$)
parts of the function $H$.
Notice that the non--resonant part of a function
that only depends on the action variable is zero.

Given a Hamiltonian $H$, we now choose $\UU_H$ to be
the canonical transformation $U_\phi$ whose generating function
$\phi$ is defined implicitly by the equation
$$
\Id^{-}\bigl(H\circ U_\phi)=0\,,
\qquad \Id^{+}\phi=0\,,
\equation(UUHDef1)
$$
provided that a (unique) solution exists.
This definition of $\UU_H$ is similar in spirit to the one used in [\rKi],
but it is considerably simpler.
In particular, if $H$ is analytic and sufficiently close
to $H_0$ (in an appropriate sense),
then equation \equ(UUHDef1) can be solved
by a relatively simple Newton type iteration [\rAK].
Concerning our definition of resonant modes,
we note that there are other possible choices that should lead
to the same results.
But if $I^{+}$ is too small (e.g. $\kappa=0$),
then the equation \equ(UUHDef1) becomes singular,
and if $I^{+}$ is chosen too large, then $\RG$ will not be
analyticity improving.

In the remaining part of this paper,
we present some numerical evidence for the existence
of a nontrivial fixed point of $\RG$
and discuss some of the expected properties of this fixed point.

\section Results

In order to simplify the problem of finding a nontrivial
fixed point for $\RG$, we will restrict our search
to a suitable subclass of Hamiltonians that is left invariant by $\RG$.
In particular, we will only consider Hamiltonians that are even in $q$, 
or equivalently, invariant under
the reflection $\Gamma: (q,p)\mapsto(-q,p)$.
The equations \equ(GenFun1) and \equ(UUHDef1) will be restricted
accordingly, to generating functions $\phi$ that are odd in $q$,
and to canonical transformations $\UU_H=U_\phi$
that commute with $\Gamma$.
We note that these parity properties of $\phi$ and $U_\phi$
follow from $H$ being even in $q$, if $H$ is close to $H_0\,$.
As a further restriction, we assume that
$$
H(q,p)=\omega\cdot p+h(q,\Omega\cdot p)
=\omega\cdot p+\sum_{\nu, n}
h_{\nu,n}\cos(\nu\cdot q)(\Omega\cdot p)^n\,,
\equation(HRestrict)
$$
with $h_{\nu,n}\in\real$.
For the generating function of the canonical transformation
$\UU_H=U_\phi$ we make the corresponding ansatz
$\phi(q,p)=\varphi(q,\Omega\cdot p)$, where $\varphi(q,z)$ is
a sum of terms $\varphi_{\nu,n}\sin(\nu\cdot q)z^n$.
Given such a function $\varphi$,
we can turn the second part of \equ(GenFun1) into an equation
for a function $Z=\Omega\cdot P$,
that only depends on $q$ and $z=\Omega\cdot p$ and is even in $q$:
$$
Z(q,z)=-\Omega\cdot\nabla_1\varphi\bigl(q,z+Z(q,z)\bigr)\,.
\equation(GenFun2)
$$
From a solution of this equation we obtain
$$
U_\phi(q,p)
=\Bigl(q+\partial_2\varphi\bigl(q,z+Z(q,z)\bigr)\Omega\,\hbox{\bf ,}\,
p-\nabla_1\varphi\bigl(q,z+Z(q,z)\bigr)\Bigr)\,, \qquad z=\Omega\cdot p\,,
\equation(GenFun3)
$$
where $\partial_2\varphi$ denotes the partial derivative
of $\varphi$ with respect to the second argument.
It is easy to check that if $H$ is of the form \equ(HRestrict),
then the same is true (formally) for $H\circ U_\phi\,$.

Within this restricted class of Hamiltonians,
the transformation $\RG$ involves the following steps:
Given $H$, we first solve the equation $\Id^{-}(H\circ U_\phi)=0$
for a generating function $\phi=\Id^{-}\phi$ of the type described above.
After composing $H$ with the canonical transformation $\UU_H=U_\phi\,$,
the new Hamiltonian is again of the form \equ(HRestrict),
and made up entirely of resonant modes.
Some of these resonant modes are turned into non--resonant ones
in the next step: $H\mapsto H\circ T_1\,$.
After this step, we perform the scaling
$H\mapsto\tau\mu^{-1}H(.,\mu .)-\epsilon$ described after \equ(RGDef1).
The resulting function $\RG(H)$ belongs again to our restricted
class of Hamiltonians.
We note that $\tau(H)=\vartheta$ for all Hamiltonians $H$ in this class,
i.e., the renormalization of time is trivial.

In our numerical implementation of these steps,
we approximate functions like $h(q,z)$ or $\phi(q,z)$
by finite  linear combinations of modes $\cos(\nu\cdot q)z^n$
or $\sin(\nu\cdot q)z^n$, respectively,
and represent them by arrays of Fourier--Taylor coefficients.
The cutoff is taken to be of the form $\|\nu\|+n\le N$,
and all intermediate results
are truncated to the same fixed ``degree'' $N$.
We have written subroutines for the addition, differentiation,
multiplication, and composition of such functions;
and similarly for the inverses of these operations,
using Newton's method when necessary.
We also use Newton's method to solve \equ(UUHDef1).
The projections $\Id^{\pm}$ are fixed once and for all,
by choosing $\sigma=2/5$ and $\kappa=\sigma^2$.

In order to find a good approximation
for the expected nontrivial fixed point $H_\ast$ of $\RG$,
we started with some initial guess,
and iterated (the numerical implementation of)
a contraction that has the same fixed point as $\RG$.
The contraction was chosen of the form $\RG\circ\PP$,
with $\PP$ a projection onto a codimension two manifold
passing through $H_\ast\,$, defined by the equation $p(\RG(H))=p(H)$,
for a suitably $\real^2$--valued function $p$.
For the degrees $N=7,9,11,13,15$, we iterated the contraction,
until reaching a Hamiltonian
$H_\ast^{\sss(N)}(q,p)=\omega\cdot p+h_\ast^{\sss(N)}(q,\Omega\cdot p)$
that satisfies the fixed point equation for $\RG_{\rm numeric}^{\sss(N)}\,$,
up to an error less than $3\times 10^{-15}$ in the norm
$$
\|h\|=\sup_{\nu,n}
|h_{\nu,n}|e^{2\|\nu\|}\,.
\equation(NormDef)
$$
At degree $N=7$, the initial guess was obtained
by a bisection procedure, as described e.g. in [\rCGJi--\rCGJK].
Then $H_\ast^{\sss(7)}$ was used as a starting point at degree $9$,
and similarly for higher degrees.
As a result of this iteration procedure, we get
approximate values for the Fourier--Taylor coefficients of $H_\ast\,$,
and for the scaling parameter $\mu(H_\ast)$.
In particular, the computed coefficients satisfy
$\|h_\ast^{\sss(N)}\|=0.92\ldots$ for all $N$, indicating
that $H_\ast$ has a decent domain of analyticity.
Our results for $\mu(H_\ast^{\sss(N)})$ are listed
in the second column of Table 1.
The $N=15$ value for the energy scaling factor $\vartheta/\mu$
differs by less than $10^{-7}$ from the value $7.0208826$ 
obtained from a renormalization group analysis
of area--preserving maps [\rKad--\rMcKiii].

The fact that $\RG\circ\PP$ was observed to be a contraction,
indicates that in the space of Hamiltonians considered,
$\RG$ has (at most) two unstable directions at $H_\ast\,$.
In order to determine these directions,
we computed the image of a two dimensional subspace
under iteration of a numerical approximation to $D\RG(H_\ast)$.
From the action of this map on the limiting ``invariant'' subspace,
we computed  approximate values for the two expanding
eigenvalues $\delta_1$ and $\delta_2$ of $D\RG(H_\ast)$;
see columns 3 and 4 of Table 1.
The value of $\delta_2$ obtained in [\rMcKiii] is $1.6279500$.
As we mentioned in the introduction, one of the eigenvalues $\delta_j$
is associated with an irrelevant degree of freedom
corresponding to a translation in the action variable.
Indeed, our numerical results are compatible with the
expected relation $\delta_1\mu=-1/\vartheta$,
involving the scaling parameter $\mu$ for the action variables.
We note that the limiting discrepancy of about $2\times 10^{-7}$
is due to the fact that we used a fixed difference quotient
for our computation, instead of writing a program for the
derivative of $\RG$.

\bigskip\bigskip

$$
\vbox{\tabskip=0pt\offinterlineskip\halign to205.8pt{
\strut#&\vrule#\tabskip=0.5em&\hfil#&\vrule#&\hfil#&\vrule#&\hfil#
&\vrule#&\hfil#&\vrule#\tabskip=0pt\cr
\noalign{\hrule}
& &$N$\hfil & &$\mu$\hfil
& &$\delta_1+1/(\mu\vartheta)$\hfil & &$\delta_2$\hfil&\cr
\noalign{\hrule}
& & 7& &$0.230376748$& &$1.1\times 10^{-5}$& &1.66289337&\cr
\noalign{\hrule}
& & 9& &$0.230454722$& &$6.8\times 10^{-7}$& &1.62803658&\cr
\noalign{\hrule}
& &11& &$0.230460077$& &$1.9\times 10^{-7}$& &1.62795826&\cr
\noalign{\hrule}
& &13& &$0.230460180$& &$2.1\times 10^{-7}$& &1.62795515&\cr
\noalign{\hrule}
& &15& &$0.230460194$& &$2.1\times 10^{-7}$& &1.62795473&\cr
\noalign{\hrule}}}
$$
\vskip-0.5em
\cl{{\small Table 1. Scaling $\scriptstyle\mu$,
and eigenvalues $\scriptstyle\delta_j$ of $\scriptstyle D\RG(H_\ast)$}}

\bigskip\bigskip

We do not expect $D\RG(H_\ast)$ to have additional eigenvalues
of modulus $\ge 1$, when considered on a space of general
Hamiltonians of the form \equ(FTDef1).
Support for this comes from earlier numerical work [\rKii],
where a renormalization group transformation similar to $\RG$
was implemented, without the restrictions
considered in this section, except for the condition
that $H(q,p)$ be an even function of $q$.
At the approximate nontrivial fixed point of this transformation,
no eigenvalues of modulus $\ge 1$ were found,
besides $\delta_1$ and $\delta_2\,$.

Next, we consider the scaling transformation $\SS_H=\UU_H\circ T_{\mu(H)}$
which maps orbits for $\RG(H)$ to orbits for $H$.
We will assume that $\RG$ has a fixed point $H_\ast$
of the form \equ(HRestrict), close to our
Hamiltonians $H_\ast^{\sss(N)}$.
Since $\SS_{H_\ast}$ commutes with the reflection $\Gamma$
and maps the energy surface $H_\ast(q,p)=E$ to the surface
of of constant energy $(E+\eps)\mu/\vartheta$,
with $|\mu/\vartheta|<1$, we expect that
$\SS_{H_\ast}$ has a fixed point of the form $(0,p_\infty)$.
Our numerical results indicate that this is the case,
and that $p_\infty\approx(0.05616862,0.01352928)$.
Since $\UU_{H_\ast}\circ T_1$ is a canonical transformation
and commutes with $\Gamma$, the derivative of $\SS_{H_\ast}$
at the point $(0,p_\infty)$ has to be of the form
$$
D\SS_{H_\ast}(0,p_\infty)=\left[A~0\atop 0~B\right]\,,
\qquad A^{\sss T}B=\mu\id\,,
\equation(Scaling)
$$
where $A$ and $B$ are $2\times 2$ matrices.
By comparing derivatives of $\RG(H_\ast)$ and $H_\ast$
at the point $(0,p_\infty)$, one also finds that
$A$ has two eigenvalues $\lambda_1=\vartheta$ and $\lambda_2\,$,
with eigenvectors $v_1=\nabla_2 H_\ast(0,p_\infty)$
and $v_2=\Omega$, respectively.
Thus, since $A^{\sss T}B=\mu\id$, the eigenvalues of $B$
are $\lambda_3=\mu/\lambda_2$ and $\lambda_4=\mu/\vartheta$,
and the corresponding eigenvectors are $v_3$ and $v_4=\omega$,
with $v_3$ perpendicular to $v_1\,$.
We note that the triviality of $\lambda_1$ and $v_4$
is related to the fact that the geometry
in the direction of the flow is trivial,
and that the vector field for $H_\ast$ is invariant
under translations $(q,p)\mapsto(q,p-s\omega)$.
The approximate values for $\lambda_2\,$, obtained from our
approximate fixed points $H_\ast^{\sss(N)}$,
are listed in column 3 of Table 2.
At $N=15$, our value for $1/\lambda_2\,$,
when truncated to $8$ digits,
agrees with the value found in [\rMcKiii] for area--preserving maps.

\bigskip\bigskip

$$
\vbox{\tabskip=0pt\offinterlineskip\halign to280.4pt{
\strut#&\vrule#\tabskip=0.5em&\hfil#&\vrule#&\hfil#&\vrule#&\hfil#
&\vrule#&\hfil#&\vrule#&\hfil#&\vrule#\tabskip=0pt\cr
\noalign{\hrule}
& &$N$\hfil & &$\lambda_1-\vartheta$\hfil & &$\lambda_2$\hfil
& &$\lambda_3-\mu/\lambda_2$\hfil & &$\lambda_4-\mu/\vartheta$\hfil &\cr
\noalign{\hrule}
& & 7& &$1.5\times 10^{-16}$& &$-0.7074647886$& &$1.2\times 10^{-4}$& &$5.2\times 10^{-5}$&\cr
\noalign{\hrule}
& & 9& &$1.5\times 10^{-16}$& &$-0.7068366347$& &$7.8\times 10^{-6}$& &$3.4\times 10^{-6}$&\cr
\noalign{\hrule}
& &11& &$1.5\times 10^{-16}$& &$-0.7067965675$& &$1.7\times 10^{-7}$& &$7.2\times 10^{-8}$&\cr
\noalign{\hrule}
& &13& &$1.5\times 10^{-16}$& &$-0.7067957946$& &$2.0\times 10^{-8}$& &$8.7\times 10^{-9}$&\cr
\noalign{\hrule}
& &15& &$1.5\times 10^{-16}$& &$-0.7067956886$& &$4.5\times 10^{-11}$& &$1.6\times 10^{-17}$&\cr
\noalign{\hrule}}}
$$
\hskip-0.5em
\cl{\small Table 2. Scaling parameters $\scriptstyle\lambda_j$}

\bigskip\bigskip

In order to visualize the critical scaling $\SS_{H_\ast}\,$,
we computed some trajectories for the reduced flow
in the variables $q$ and $z=\Omega\cdot p$,
$$
(\dot q,\dot z)=\bigl(\omega+\partial_2 h(q,z)\Omega\,\hbox{\bf ,}
-\Omega\cdot\nabla_1h(q,z)\bigr)\,,
\equation(ZFlow)
$$
for the Hamiltonian $H_\ast^{\sss(15)}\,$.
The reduced phase space is considered to be $\torus^2\times\real$,
that is, $q$ is identified with $q+2\pi\integer^2$.
Figure 1 shows $14$ orbits of the return map
for a rectangle in the plane $\Pi=\{(sv,z): s,z\in\real\}$,
where $v=(-0.899,1)$.
The coordinates in the horizontal and vertical direction
are $v\cdot q$ and $z$, respectively. In these coordinates,
the scaling fixed point is at $(0,z_\infty)$,
where $z_\infty=\Omega\cdot p_\infty=0.0478\ldots$.
The curve $\CC$ through this point represents the intersection
of the critical $\omega$--torus with $\Pi$.
We note that $v$ is not an eigenvector of $A$.
The vector $v$ has been chosen in such a way that $\CC$
has zero curvature at the point $(0,z_\infty)$.
In order words, we used the flow
to linearize $\SS_{H_\ast}$ approximately, in the angle variable.
Figure 2, which will be discussed later,
indicates that for most other choices of $v$, $\CC$
would have a nonzero curvature at $q=0$.
As a result, the vertical scaling off the line $v\cdot q=0$
would be dominated by $\lambda_2^2$ instead of $\lambda_3\,$.
The points in Figure 1, that lie on the symmetry line $v\cdot q=0$
and are surrounded by circles,
belong to periodic orbits with periods $\vartheta_5=13/8$,
$\vartheta_6=21/13$, $\ldots$.
If a similar picture holds for the first few $\vartheta_k$--orbits
of $H_\ast\,$, then the corresponding points $(0,z_k)$
have to accumulate at $(0,z_\infty)$, with an asymptotic ratio
$(z_{k+1}-z_\infty)/(z_k-z_\infty)\to\lambda_3=-0.326\ldots$.
And in the direction $(v,0)$, the orbits have to scale
with an asymptotic ratio equal to $\lambda_2\,$.
These properties follow from the fact that $H_\ast\circ\SS_{H_\ast}$
is a constant multiple of $H_\ast\,$, which implies
(since $\SS_{H_\ast}$ is canonical modulo a scaling of the actions)
that if an orbit for $H_\ast$ is periodic with frequency vector $\nu$,
then its image under $\SS_{H_\ast}$ is another periodic orbit
for $H_\ast\,$, with frequency vector $T^{-1}\nu$.
The observed accumulation of $\vartheta_k$--orbits also suggests
that the point $(0,p_\infty)$ lies on an invariant $\omega$--torus
for $H_\ast\,$,
and that this torus is mapped onto itself by $\SS_{H_\ast}\,$.
For corresponding results on area--preserving map,
we refer to [\rKad--\rMcKiii].

The results obtained here and in [\rKii] suggest that,
as far as the breakup of smooth $\omega$--tori is concerned,
the unstable manifold of the transformation $\RR$
described after \equ(GenFun1)
represents the ``typical'' one--parameter family of Hamiltonians.
On the other hand,
these Hamiltonians are all of the form \equ(HRestrict).
Their $\omega$--tori at different energies have to break up simultaneously,
since they are just translates of each other.
In order to see in what sense this may be typical,
consider some ``arbitrary'' Hamiltonian $H(q,p)$
that is even in $q$ and of the form \equ(FTDef1).
Suppose that $H$ allows us to define
a differentiable one--parameter family of area--preserving maps 
$E\mapsto\Psi_E$, by restricting the flow for $H$
to surfaces $H(q,p)=E$ and choosing an appropriate Poincar\'e section.
Assume now that $H$ has a critical $\omega$--torus
for an energy $E_\infty\,$, and that the family
$E\mapsto\Psi_E$ intersects the critical manifold $\WW$
for the area--preserving renormalization group fixed point
transversally, at $E=E_\infty\,$.
Then, according to the renormalization group picture
for area--preserving maps, there is an asymptotic scaling
that relates the orbits of $\Psi_E$ to orbits of $\Psi_{f(E)}\,$,
where $f(E_\infty+\eps)=E_\infty+\delta_2\eps+\OO(\eps^2)$.
Since these orbits all stem from a single Hamiltonian $H$,
just considered on different energy surfaces,
one would expect to find an asymptotic scaling rate
$\delta_2^{-1}=0.614\ldots$ in the phase space of $H$,
in a direction transversal to the surface $H(q,p)=E_\infty\,$.
But our results yield a scaling rate $\lambda_4=0.142\ldots$
in this direction.
This suggests that the family $E\mapsto\Psi_E$
cannot be transversal to $\WW$.
In addition, if $E\mapsto\Psi_E$ and $\WW$ are smooth, then
the fact that $\delta_2^{-1}$ is not an integer power
of $\lambda_4$, seems to imply that the distance between $\Psi_E$
and its projection onto $\WW$ tends to zero faster
than any power of $E-E_\infty\,$, as $E$ approaches $E_\infty\,$.
In the analytic case,
this would lead to the conclusion that $H$ has either no
critical torus, or else an entire family of critical tori.
We have not done any numerical investigations
to confirm such behavior, or to find a counter--example.

We conclude with some observations concerning the
critical $\omega$--torus for the fixed point $H_\ast\,$,
on the energy surface containing the point $(0,p_\infty)$.
A numerical approximation of this torus, in the
coordinates of the reduced flow \equ(ZFlow), is shown in Figure 2.
These data suggest e.g. that the critical torus of $H_\ast$
is the graph of a differentiable function $p=G(q)$,
with $G(0)=p_\infty\,$, and that $0$ is a ``saddle point'' of
the corresponding function $z=\Omega\cdot G(q)$.
Notice that by symmetry,
$\Omega\cdot G$ has to have a critical point at $q=0$.
The fact that the unstable eigenvector
of $D\SS_{H_\ast}(0,p_\infty)$ is tangent to the flow at $(0,p_\infty)$,
suggests that the orbit of the point $(0,p_\infty)$
coincides with the unstable manifold of $\SS_{H_\ast}$ at this point,
and this in turn suggests that the critical $\omega$--torus
is an attractor for the scaling map $\SS_{H_\ast}\,$.
In order to obtain information about the regularity
of this torus, it is useful to look at periodic points of $\SS_{H_\ast}\,$.
Consider e.g. the fixed point $(0,p_\ast)$.
Our earlier discussion of $\SS_{H_\ast}$ suggests
that the degree of differentiability of the critical $\omega$--torus
at this point is no more than $\ln|\lambda_3|/\ln|\lambda_2|=3.229\ldots$.
The same holds of course for every point on the orbit of $(0,p_\ast)$
under the flow defined by $H_\ast\,$.
But there are other (dense sets of) points where
the degree of differentiability is expected to be less than $2$.
Consider e.g. the set of points on $\torus^2$
that are invariant under the reflection $q\mapsto-q$.
This set consists of $\gamma_0=(0,0)$, and three other points
$\gamma_1,\gamma_2,\gamma_3$,
whose coordinates are both either $0$ or $\pi$, modulo $2\pi$.
Notice that, according to Figure 2,
these are precisely the critical points of $\Omega\cdot G$.
It is easy to see that $\UU_{H_\ast}$ leaves the planes
$\Gamma_i=\{\gamma_i\}\times\real^2$ invariant.
Thus, since $T$ acts as a permutation
on $\{\gamma_1,\gamma_2,\gamma_3\}$,
the corresponding planes $\Gamma_i$ are permuted by $\SS_{H_\ast}\,$.
Assuming that the critical $\omega$--torus
is invariant under $\SS_{H_\ast}\,$, this implies that
$\{G(\gamma_1),G(\gamma_2),G(\gamma_3)\}$
is an orbit of period $3$ for $\SS_{H_\ast}\,$.
An estimate on the degree of differentiability
of the critical torus at $G(\gamma_1)$
can now be obtained by linearizing $\SS_{H_\ast}^3$ at this point,
and then computing the analogue of the ratio $\ln|\lambda_3|/\ln|\lambda_2|$.
Based on observations for area--preserving maps [\rSK,\rMcKi],
we expect the value of this ratio to be $1.790\ldots$,
but we did not verify this numerically.
We note that for every periodic orbit of $T$ on the torus $\torus^2$,
there is a corresponding periodic orbit of the transformation $\RG$.
This is a consequence of the following property of $\RG$:
Assuming that the equation \equ(UUHDef1) has a unique solution,
it is easy to verify that $\RG\circ\JJ_\gamma=\JJ_{T^{-1}\gamma}\circ\RG$,
where $\JJ_\gamma$ denotes the rotation $H\mapsto H(.-\gamma,.)$.

\section Conclusion

The renormalization group transformation $\RG$ introduced in this paper
can be used to study, and possibly explain, the breakup of golden
invariant tori in Hamiltonian systems with two degrees of freedom.
We find an approximate numerical fixed point for $\RG$,
whose critical indices and scaling parameters
are in excellent agreement with the corresponding
values for area--preserving maps.
The same methods are expected to work for other
reduced quadratic irrationals, with similar results.
Two distinguishing features of our transformation are
that it applies to Hamiltonians of a general type,
and that it seems to be analyticity improving.
On a conceptual level, this indicates that
all irrelevant degrees of freedom are being eliminated
under the iteration of $\RG$.
Numerically, the analyticity improving property allows for high accuracy,
as truncation errors decrease exponentially
in the degree of the truncation.
Our results not only indicate that Hamiltonians with critical
invariant $\omega$--tori are asymptotically self--similar,
but they also provide detailed geometric information
about the relevant scalings and normal forms.
Further investigations will be necessary
in order to determine the implications of these findings
on the mechanisms underlying the breakup of invariant tori.
Another project for future research
is to carry out a computer--assisted proof
for some of the results presented in this paper.

\bigskip\noindent{\bf Acknowledgments}\hfill\break
We would like to thank R.~de~la~Llave for helpful discussions.

\references

\baselineskip=11.8pt plus 2pt

\item{[\rED]} D.F.~Escande, F.~Doveil, {\it
Renormalisation Method for Computing the Threshold of the Large Scale
Stochastic Instability in Two Degree of Freedom Hamiltonian Systems.}
J. Stat. Phys., {\bf 26}, 257--284 (1981).

\item{[\rMEsc]} A.~Mehr, D.F.~Escande, {\it
Destruction of KAM Tori in Hamiltonian Systems:
Link with the Destabilization of nearby Cycles
and Calculation of Residues.}
Physica, {\bf 13D}, 302--338 (1984).

\item{[\rMcKii]} R.S.~MacKay, {\it
Three Topics in Hamiltonian Dynamics.}
Preprint U. Warwick (1994).\hfil\break
Also in: ``Dynamical Systems and Chaos'', Vol.2,
Y.~Aizawa, S.~Saito, K.~Shiraiwa (eds),
World Scientific, London (1995).

\item{[\rKi]} H.~Koch, {\it
A Renormalization Group for Hamiltonians,
with Applications to KAM Tori.}
Preprint U. Texas, mp\_arc 96--383 (1996),
to appear in Erg. Theor. Dyn. Syst.

\item{[\rKii]} H.~Koch, {\it
unpublished numerical results}
(1997).

\item{[\rCGJi]} C.~Chandre, M.~Govin, H.R.~Jauslin, {\it
KAM--Renormalization Group Analysis of Stability in Hamiltonian Flows.}
Phys. Rev. Lett. {\bf 79}, 3881--3884 (1997).

\item{[\rCGJii]} M.~Govin, C.~Chandre, H.R.~Jauslin, {\it
KAM--Renormalization Group Approach to the Breakup
of Invariant Tori in Hamiltonian Systems.}
Phys. Rev. E {\bf 57}, 1536--1543 (1998).

\item{[\rCGJK]} C.~Chandre, M.~Govin, H.R.~Jauslin, H.~Koch, {\it
Universality for the Breakup of Invariant Tori in Hamiltonian Flows.}
Preprint U. Bourgogne, mp\_arc 98--51 (1998),
to appear in Phys. Rev. E.

\item{[\rAK]} J.J.~Abad, H.~Koch {\it
work in progress.}


\item{[\rKad]}
L.P.~Kadanoff, {\it
Scaling for a Critical Kolmogorov--Arnold--Moser Trajectory.}
Phys. Rev. Lett. {\bf 47}, 1641--1643 (1981).

\item{[\rSK]} S.J.~Shenker,  L.P.~Kadanoff {\it
Critical Behavior of a KAM Surface. I. Empirical Results.}
J. Stat. Phys. {\bf 27}, 631--656 (1982).

\item{[\rMcKi]} R.S.~MacKay, {\it
Renormalisation in Area Preserving Maps.}
Thesis, Princeton (1982).
World Scientific, London (1993).

\item{[\rMcKiii]} R.S.~MacKay, {\it
A Renormalization Approach to Invariant Circles
in Area--Preserving Maps.}
Physica D {\bf 7}, 283--300 (1983).

\item{[\rSti]} A.~Stirnemann, {\it
Towards an Existence Proof of MacKay's Fixed Point.}
Comm. Math. Phys. {\bf 188}, 723--735 (1997).


\item{[\rMH]} J.~M. Mao and R.~H.~G. Helleman, {\it
Existence of KAM Tori for Four--Dimensional
Volume Preserving Maps.}
Nuovo Cimento {\bf 104}B, 177--182 (1989).

\item{[\rACS]} R.~Artuso, G.~Casati, D.L.~Shepelyansky, {\it
Breakdown of Universality in Renormalization Dynamics
for Critical Invariant Torus.}
Europhys. Lett. {\bf 15}, 381--386 (1991).

\item{[\rMMS]} R.S.~MacKay, J.D.~Meiss, J.~Stark, {\it
An Approximate Renormalization for the Break-up
of Invariant Tori with Three Frequencies.}
Phys. Lett. A {\bf 190}, 417--424 (1994).


\item{[\rTh]} W.~Thirring, {\it
A Course in Mathematical Physics I: Classical Dynamical Systems.}
Springer Verlag, New York $\cdot$ Wien (1978).

\item{[\rRdL]} R.~de~la~Llave, {\it
Introduction to KAM Theory.}
Preprint U. Texas, mp\_arc 93--8 (1993).


\vfil\eject
\ipsfig 8truein; 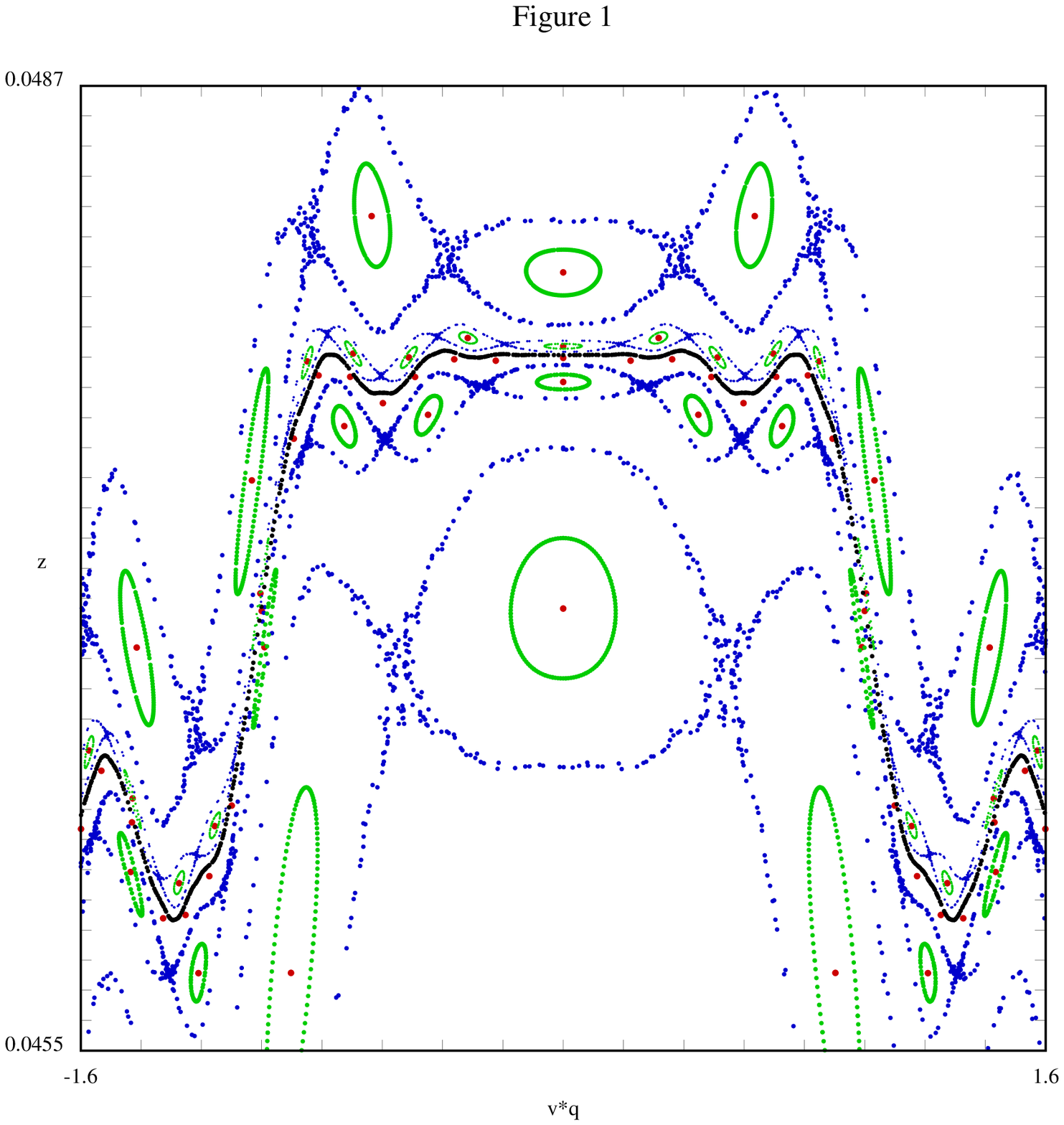;
\vfill\eject
\ipsfig 7truein; 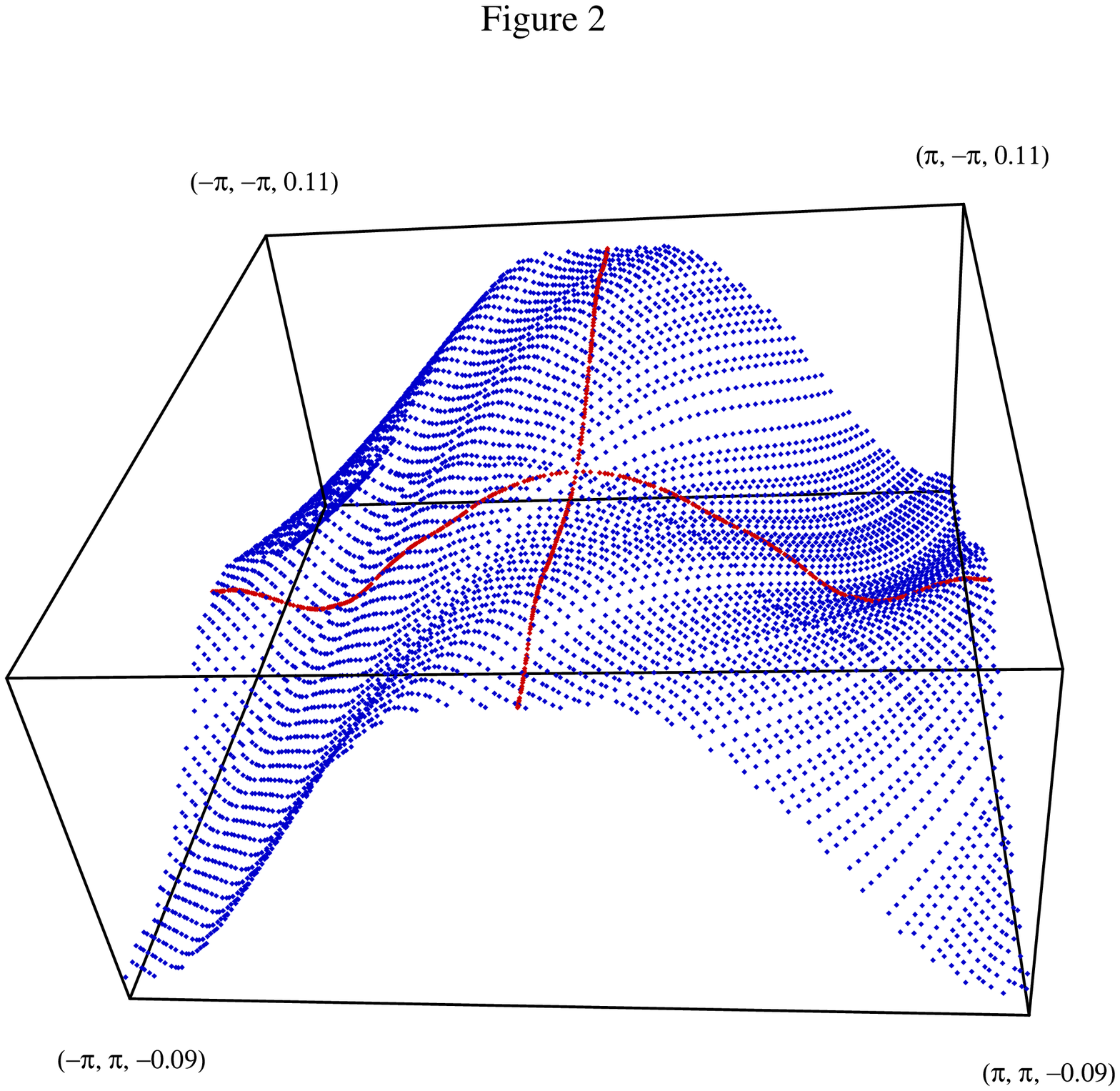;
\bye